# Hamilton's equations of motion from the SU(1,1)/U(1) symmetry of Schwarzschild radial dynamics

S. H. Castles[*]
*Mosier, OR*

(Dated: February 11, 2011)

Particles moving on a radial ray with respect to a Schwarzschild mass are shown to have SU(1,1)/U(1) dynamical symmetry. This symmetry is used to identify a global time variable shared by all test particles moving on a radial ray. With this time variable, Hamilton's equations for particles on a radial ray have the form of the harmonic oscillator. The SU(1,1)/U(1) symmetry, in conjunction with the well defined time variable and observer, assists in determining the quantization of the motion of the test particles.

## I. INTRODUCTION

In a seminal paper using string theory, the symmetry group of the equations of motion for a particle moving on a radial ray with respect to a Schwarzschild mass was deduced to be SU(1/1)/U(1)[1]. Herein it is demonstrated that the dynamical symmetry group for the family of all particles moving on such a radial ray is SU(1,1)/U(1). Using the SU(1,1)/U(1) symmetry, it is shown that this family of particles shares a global time parameter. The global time parameter is related to the proper time of each particle which, in usual manner, is related to the Schwarzschild time parameter, the time measured by an observer at rest on the radial ray and in the limit of infinite distance relative to the Schwarzschild mass.

The use of the symmetry group precisely defines the time variables and the role of the observer. An incomplete definition of a global time parameter and of the role of the observer has interfered with the quantization of general relativistic geometries using quantum field theoretical approaches.[2]

Using the global time parameter, a set of canonical variables is found that yields Hamilton's equations in harmonic oscillator form. The SU(1,1)/U(1) dynamical symmetry of this family of particles assists in the clarification of the quantization of the motion.[3,4]

In Section 2 the geodesics in the geometry corresponding to the SU(1,1)/U(1) symmetry are described relative to two models of hyperbolic geometry. This is accomplished by defining a pair of Möbius transformations that generates the geodesics. The geodesics are related to the geodesics of the Poincaré disk model of hyperbolic geometry by a rotation about the origin. This geometry is then converted to a portion of the Poincaré upper half-plane model.

In Section 3 the family of all particles on the radial ray is delineated. As the family of particles is delineated, their geodesics are related to the geodesics within the Poincaré disk model. The initial conditions determining the motion of each particle on the radial ray are related to a set of arcs of circles that are everywhere orthogonal to the geodesics. Then the radial equation of motion for the family of particles are related to the same geodesics and arcs as they appear in the Poincaré upper-half plane model of hyperbolic geometry. In Section 4 it is noted that the two first order equations of motion are not independent in the two dimensional geometry of the radial ray. In Section 5 a set of canonical variables are defined that yield Hamilton's equations of motion in quadratic form.

[*]E-mail: shcastles@embarqmail.com

Two appendices are included. In Appendix 1 a brief introduction to projective coordinates for SU(1,1)/U(1) is presented[5]. In Appendix 2, the geometry of the SU(1,1)/U(1) symmetry group is used to introduce an analytic continuation across the Schwarzschild horizon.

Note that the results of Section 3 are independent of the method used to determine the desired SU(1,1)/U(1) parameterization of the physical quantities. Similarly, once the parameterization of the physical quantities required to identify Hamilton's equations of motion are known, the resulting Hamiltonian and canonical position and momentum are valid independent of the method used to derive them.

**II. COSET SPACE GEOMETRY**

The points in the SU(1,1)/U(1) coset space, denoted by z, can be defined on the open unit disk in the complex plane whose boundary is the unit circle. (See Appendix 1 for an introduction to projective coset space geometry.[5]) A common parameterization of the points in the unit disk is given by

$$z = \tanh(\chi)(\hat{x} + i\hat{y}). \tag{1}$$

with $\hat{x}^2 + \hat{y}^2 = 1$ and $-\infty < \chi < \infty$. Often, as an alternative parameterization of the points in the unit disk, z is defined by a transformation in the radial direction times a rotation, $z = \tanh(\chi) \circ \exp(i\theta)$. The flow lines[6] corresponding to these two transformations are two orthogonal families of curves, a family of radial lines through the origin which are the flow lines of the radial transformation and a family of circles centered on the origin which are the flow lines of the rotation.

Another parameterization of the points in the unit disk is represented geometrically by the Poincaré disk model of hyperbolic geometry[7]. Transformations (2a) and (2b) below produce a family of geodesics with SU(1,1)/U(1) symmetry. The geodesics are related to the geodesics within the Poincaré disk model. Specifically, if the geodesics in the geometry with SU(1,1)/U(1) symmetry are allowed an additional degree of freedom, namely a rotation about the origin of the disk, the resulting geodesics are the family of geodesics in the Poincaré disk model.

Define Möbius transformations

$$z \mapsto \frac{\cosh(\chi_1)z + \sinh(\chi_1)}{-\sinh(\chi_1)z + \cosh(\chi_1)} \tag{2a}$$

$$z \mapsto \frac{\cosh(\chi_2)z + i\sinh(\chi_2)}{-i\sinh(\chi_2)z + \cosh(\chi_2)} \tag{2b}$$

with $-\infty < \chi_1 < \infty$ and $-\infty < \chi_2 < \infty$. Transformation (2a) will be called the "geodesic transformation" since the flow lines of this transformation are geodesics within the geometry of the SU(1,1)/U(1) coset space. (They are also geodesics within the Poincaré disk model.) Transformation (2b) will be called the "extremum transformation" for physical reasons explained in Section 3.

For each point in the unit disk, the flow lines of the geodesic transformation, (2a), define arcs of circles. Each arc is centered on the y-axis and is orthogonal to the unit circle. (See Fig. 1, in which z = x + iy.) The flow lines of the geodesic transformation will be shown to be the geodesics of particles moving on a radial ray relative to a Schwarzschild mass.

The flow lines of the extremum transformation are arcs of circles centered on the x-axis. The arcs have two fixed points, $z = \pm i$ or, with $z = x + iy$, the points $(x,y) = (0, \pm 1)$. The tangent to the arcs at the two points preserved by the extremum transformation makes an angle of magnitude $\varepsilon$ with respect to the y-axis (see Fig. 1), where $\varepsilon$ is related to $\chi_1$ by the following useful identities: $\tan(\varepsilon/2) = \tanh(2\chi_1)$, $\tan(\varepsilon) = \sinh(2\chi_1)$, $\cos(\varepsilon) = \text{sech}(2\chi_1)$, and in differential form by

$$d\varepsilon = \text{sech}(2\chi_1) d(2\chi_1) \quad (3)$$

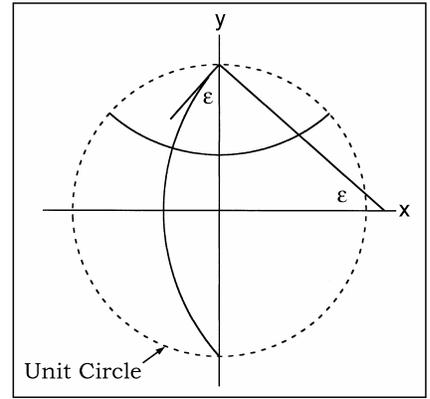

Fig. 1: The solid arc centered on the y-axis is the flow line of the geodesic transformation, Eq. (2a), acting on the point $(x,y) = (0, \tanh(0.4))$ or, equivalently, an extremum transformation, Eq. (2b), of the x-axis, $|x| < 1$. The solid arc centered on the x-axis is the geodesic transformation with $\chi_1 = -0.4$ of the y-axis, $|y| < 1$. It makes an angle $\varepsilon$ relative to the y-axis at $y = \pm 1$.

The angle $\varepsilon$ can serve to parameterize the points on the arcs of the circles centered on the y-axis, the flow lines of the geodesic transformation. Note that the geodesic transformation, acting on the points on the y-axis, $-1 < y < 1$, transform the y-axis into points on the arc of a circle centered on the x-axis, the arc being the flow lines of an extremum transformation. These arcs can be characterized by the value of the intercept of the arc with the x-axis, specifically

$x = \tanh(\chi_1)$. Similarly, the extremum transformation, acting on the points on the x-axis, $-1 < x < 1$, transform the x-axis into points on the arc of a circle centered on the y-axis, this arc being the flow lines of a geodesic transformation.

The points in the unit disk can be defined by the product of these two transformations acting on the origin. For example, if the extremum transformation acts on the origin and the geodesic transformation acts on the result, we obtain from (2)

$$z \mapsto \frac{i\cosh(\chi_1)\tanh(\chi_2) + \sinh(\chi_1)}{-i\sinh(\chi_1)\tanh(\chi_2) + \cosh(\chi_1)} \quad (4)$$

Rationalizing the denominator and using the hyperbolic sign and cosine identities for double angles and the identities

$$\tanh(\chi) = \frac{\sinh(2\chi)}{\cosh(2\chi) + 1} \quad (5)$$

and

$$\tanh^2(\chi) = \frac{\cosh(2\chi) - 1}{\cosh(2\chi) + 1} \quad (6)$$

we obtain for the points in the unit disk

$$(x, y) = \left( \frac{\sinh(2\chi_1)}{1 + \cosh(2\chi_1)\cosh(2\chi_2)}, \frac{\cosh(2\chi_1)\sinh(2\chi_2)}{1 + \cosh(2\chi_1)\cosh(2\chi_2)} \right). \quad (7)$$

A similar calculation with these identities demonstrates that if a geodesic transformation acts on the origin and the extremum transformation acts on the result, one obtains the identical result.

The continuation of the arcs defined within the unit disk form circles that fill the complex plane The continuation of the arcs within the unit disk is relevant to the continuation of the analysis presented herein to points within the Schwarzschild horizon and is briefly discussed in Appendix 2.

We wish to compare this geometry to the radial equation of motion for particles on geodesics along a radial ray with respect to a Schwarzschild mass. For this purpose it is beneficial to transform from the above geometry, a portion of the Poincaré unit disk model, into a portion of the Poincaré upper half plane model of hyperbolic geometry[7]. The standard procedure for converting the Poincaré unit disk model into the Poincaré upper half-plane model is to use the Möbius transformation

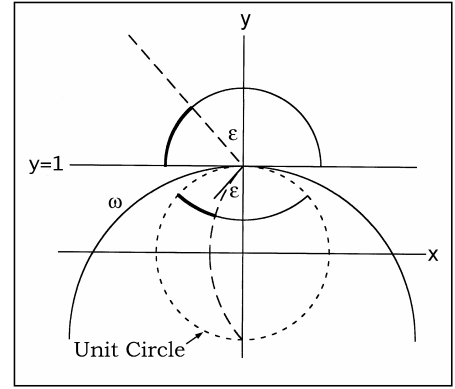

Fig. 2: The arcs from Fig. 1 are repeated and the inversion in the circle $\omega$ of the portion of the arcs with $y \geq 0$ is shown. A radial geodesic of a particle, shown for $\chi_{10} = -0.4$ and $\chi_{20} = +0.4$, is represented by the darkened portion of the arc in the unit disk. The inversion in $\omega$ of the geodesic is also shown.

$$z \mapsto z' = (iz - 1)/(-z + i) \tag{8}$$

However, in keeping with the geometric theme, an equivalent mapping will be used herein. The mapping is obtained by performing an inversion in a circle[6] on the points in the unit disk defined by Eq. (7), followed by an isometric contraction by a factor of 2. (See Fig. 2. The contraction by a factor of 2 is not included in Fig. 2 but will be accounted for analytically in (10). Also, 1 must be subtracted from the y-values to obtain the upper half-plane.) Define $\omega$ to be a circle centered at $(0,-1)$ with radius equal to 2. Perform an inversion in $\omega$ on the points z in the unit disk and label the new points by $z'$, namely

$$z \mapsto z' = -i + 4/(\bar{z} - i) . \tag{9}$$

The arcs of circles centered on the y-axis are transformed into concentric arcs of circles centered at $(0,1)$. Specifically, the arc that intercepts the y-axis at $\tanh(\chi_2)$ is transformed under the inversion in $\omega$ into an arc centered on $(0,1)$ with radius $\rho = 2\exp(-2\chi_2)$. These arcs are geodesics in the upper half-plane model. The geodesics with $SU(1,1)/U(1)$ symmetry are related to the geodesics in the Poincaré upper half-plane model by a translation along the real axis.

The arcs centered on the x-axis are transformed into lines through $(0,1)$, with the arc that intercepts the x-axis at $\tanh(\chi_1)$ transformed into a line that makes an angle $\varepsilon$ with respect to the positive y-axis. The inversion in $\omega$ transforms the unit disk to the $y > +1$ portion of the complex plane. The points in the unit disk can be defined by the intersections of the circles centered on $(0,1)$ and the lines through $(0,1)$. The inversion in the circle $\omega$ of the point defined by (7) yields the intersection point, labeled $(x', y')$, on the arc of the circle with radius $\rho$:

$$(x', y') = (\tanh(2\chi_1)\rho, 1 + \mathrm{sech}(2\chi_1)\rho) , \tag{10a}$$

$$\rho = 2\exp(-2\chi_2) . \tag{10b}$$

## III. IDENTIFICATION OF THE COSET SPACE GEOMETRY WITH THE EQUATIONS OF MOTION

The physical system considered herein is the family of test particles moving on radial geodesics along a radial ray relative to a Schwarzschild mass, m, as measured by an observer who is stationary with respect to the Schwarzschild mass, who is on the radial ray, and whose distance from the Schwarzschild mass is in the limit as the radius approaches infinity. Such an observer is often called "the observer at infinity". The complete family of particles on a radial ray will be delineated below.

The radial Schwarzschild geometry described herein is associated with a two-dimensional, radial Schwarzschild metric

$$ds^2 = -(1-2m/r)dt^2 + \frac{1}{(1-2m/r)}dr^2 \tag{11}$$

spanned by the Schwarzschild time of the observer at infinity and Schwarzschild radial coordinate, t and r. Natural units are used, c = G = 1. Normally the Schwarzschild metric is given as four dimensions, representing a four-dimensional geometry with spherical spatial symmetry. The discussion here is limited to two-dimensional, radial geometry but the spherical symmetry of the Schwarzschild mass makes this difference immaterial for radial geodesics.

The correspondence between the features of this dynamical system of particles and the SU(1,1)/U(1) coset space defined in the previous section is presented below. The elements of the correspondence are self-consistent and define a one-to-one mapping between the coset space parameters and the parameters of the equations of motion for all particles on radial Schwarzschild geodesics along a radial ray.

The origin of the unit disk, (x,y) = (0,0), corresponds to an observer (particle) at infinity and at rest relative to the Schwarzschild mass. Physically, this observer at infinity experiences a flat spacetime which can be parameterized with an orthogonal coordinate system. Geometrically, this point possesses an orthogonal coordinate system defined by the su(1,1) mod u(1) Lie algebra. The boundary of the unit disk corresponds to r = 2m, the Schwarzschild radius (the horizon). The stationary observer in the limit of infinity can only communicate with an observer (particle) outside the Schwarzschild radius, which corresponds to within the unit disk. Note that the point in the unit disk at the origin is excluded physically since communication between particles only exists for observers in the limit as the radius approaches infinity.

A particle on a radial geodesic has a set of initial conditions relative to the observer at infinity. These initial conditions are specified by the coordinates of a point in the coset space geometries of Section II. This initial condition point can be defined geometrically by performing a Möbius transformation on the origin of the coset space. The initial condition point can be specified in the unit disk geometry or in the $(x',y')$ coordinate system resulting from the inversion in the circle ω. Using (10) the initial condition point is specified by

$$(x'_0, y'_0) = (\tanh(2\chi_{10})\rho_0, 1+\text{sech}(2\chi_{10})\rho_0) \tag{12}$$

with $\rho_0 = 2\exp(-2\chi_{20})$.

The geodesic of a particle is a curve in the coset space containing the point corresponding to the initial condition point. Below, one of the known first order equations of motion defining the dynamics of a particle on a geodesic is equated with the curve

$$x'^2 + (y'-1)^2 = \rho_0^2 \tag{13}$$

where x′ and y′ are given by (10) with $\rho = \rho_0$, corresponding to $\chi_2 = \chi_{20}$. $\rho$ and $\chi_2$ are set by the initial conditions. That is, portions of the circles centered at (0,1) and defined by (10) are the geodesics for any radially moving particle whose initial conditions correspond to a point on that circle. (See Fig. 2.)

Adding translations to special relativity is normally accomplished by extending the symmetry group SO(3,1) to include translations, resulting in the Poincaré group. For the dynamics of particles on radial Schwarzschild geodesics a transformation from the origin of the coset space to an initial radial position and momentum is accomplished by a transformation that is within SU(1,1)/U(1). The dynamics along the radial Schwarzschild geodesics, as measured by the observer at infinity, are defined within the SU(1,1)/U(1) symmetry group.

In Schwarzschild coordinates the first order equations of motion for a particle moving radially with respect to a Schwarzschild mass are given by[8-11]

$$dr/d\tau = h\sqrt{E^2 - (1 - 2m/r)} \tag{14a}$$

$$dt/d\tau = \frac{E}{(1 - 2m/r)} \tag{14b}$$

where $\tau$ is the proper time per unit mass of the particle and h is +1 (-1) for an outgoing (ingoing) particle with respect to the Schwarzschild mass, m. The particles can be classified by their initial condition parameters, E, $r_0$, $\tau_0$ and h. (It will be shown that $\tau_0$ and $\varepsilon_0$ are related to one another so $\varepsilon_0$ can be used as the initial condition parameter.) For E < 1 the particle is bound. The geodesics of outgoing bound particles have a maximum radial distance from the Schwarzschild mass and ingoing bound particles would reach a maximum radial distance if the momentum of the particles were reversed. This "extremum" radius, $r_E$, is not necessarily the initial radius. For bound particles[8,9,12] $E = \sqrt{1 - 2m/r_E}$. For $E \geq 1$ the extremum condition for the particle corresponds to the particle being "at infinity" (in the limit as r approaches infinity) relative to the Schwarzschild mass, with "energy at infinity" given by the special relativistic energy per unit mass of the particle[8,12], $E = 1/\sqrt{1 - v^2}$. The velocity $v = v/c$ is the velocity the particle would have "at infinity" as measured by the (stationary) observer at infinity.

A comparison of (13) and (14a) motivates the following definitions for the parameters of the coset space in terms of the parameters of the equations of motion:

$$dr/d\tau = \tanh(2\chi_1)E \tag{15a}$$

$$\sqrt{1 - 2m/r} = \text{sech}(2\chi_1)E \tag{15b}$$

with $E = \rho_0/2 = \exp(-2\chi_{20})$. E, and therefore $\chi_2$, is a constant on any geodesic and is set by the initial conditions. The value $E < 1$, the bound particle case, corresponds to $\chi_{20} > 0$ while $E \geq 1$ corresponds to $\chi_{20} \leq 0$. With $\chi_{10} = 0$ the particle is initially at the extremum point. This extremum condition corresponds to the initial conditions being set by a pure extremum transformation - thus, the name.

The angle $\varepsilon$ has a unique value for each point on each circle centered on (0,1). (See Fig. 2.) $\varepsilon$ is thus a time-like variable for particle motion on the geodesics. Consider the invariance of $\varepsilon$. For a particular particle, $\varepsilon$ is defined geometrically as the angle relative to the angle $\varepsilon_0$ at the initial conditions point. For each particle, $\varepsilon_0$ parameterizes the initial condition point relative to the extremum point. Thus, the value of $\varepsilon$ for any two particles moving on geodesics along a particular arc of a circle (a particular value of E) differs by at most a constant set by the initial conditions.

Particles with different values of E, which are set by the initial conditions and which could correspond to a pure extremum transformation, are on different circles centered on (0,1). The extremum transformation is a Möbius transformation. All Möbius transformations are conformal and therefore preserve angles. Inversion in a circle is also a conformal transformation. Thus, particles whose initial conditions differ only by their extremum transformation have the same value of $\varepsilon$. Combining this result with the result from the previous paragraph proves that all particles with geodesics on a radial ray share the parameter $\varepsilon$ to within a constant set by the initial conditions. In this sense, $\varepsilon$ is a "global" variable.

Eqs. (15a), (15b) and (3) can be used to determine the relationship between the proper time and the parameter $\varepsilon$,

$$d\tau = -4mE(1 - E^2\cos^2(\varepsilon))^{-2}\cos(\varepsilon)d\varepsilon . \tag{16a}$$

For E < 1 this is equivalent to

$$d\tau = -(r^2/m)(1 - 2m/r)^{1/2}d\varepsilon . \tag{16b}$$

Note $d\tau/d\varepsilon$ approaches infinity as E approaches 1 and $\varepsilon$ approaches 0, the point corresponding to the stationary observer at infinity. This is consistent with a particle starting at rest at infinity. Also note that both the Schwarzschild time, which is the time measured by the observer at infinity, and the argument of the hyperbolic function, $\chi_1$, approach infinity at the Schwarzschild horizon. Thus, the coset space coordinates can not be used to describe radial geodesics inside the horizon unless an appropriate coordinate transformation is applied. Finally, from the usual expression of $d\tau$ versus $d\varepsilon$

$$d\tau = (1/E)(1 - 2m/r)dt \tag{17a}$$

and from (16b) in the form

$$d\varepsilon = -(1/E)(m/r^2)(1 - 2m/r)^{1/2}dt \tag{17b}$$

we see that while $d\tau$ changes sign inside the Schwarzschild radius, relative to dt, $d\varepsilon$ becomes imaginary inside the Schwarzschild radius.

## IV: INTERDEPENDENCE OF THE EQUATIONS OF MOTION FOR PARTICLES ON A RADIAL RAY

Only one equation of motion has been used, namely (14a) for dr/d$\tau$. Therefore it is noteworthy that in two-dimensional radial Schwarzschild geometry the two first order equations of motion and the metric are not independent. Timelike variables related to the proper time have been discussed in previous investigations of the family of radial Schwarzschild solutions[8-11]. Refs 8-11 use a four-dimensional geometry with spherical spatial symmetry. Ref. 8 notes that when two families of surfaces exist that are everywhere orthogonal and one family of surfaces consists of the geodesics, then there exists a function T such that

$$u_\alpha = -\text{ constant} * \delta_\alpha T \tag{18}$$

where $u_\alpha$ are the covariant components corresponding to $\{u^\alpha\} = \{dx^\alpha/d\tau\}$. The family of spacelike surfaces orthogonal to the geodesics is then the family of surfaces of constant T. In the two-dimensional geometry specified by the coordinates (x′, y′) this family of spacelike curves is the family of straight lines through (0,1), the curves of constant angle $\varepsilon$.

In two-dimensional radial geometry, $\{u^1, u^2\} = \{dt/d\tau, dr/d\tau\}$. With the constant in (18) equal to E,

$$u_1 u^1 + u_2 u^2 = -E(\partial_t T \frac{dt}{d\tau} + \partial_r T \frac{dr}{d\tau}) = -1. \tag{19}$$

Using both equations of motion, (14a) and (14b), (18) yields

$$dT = dt - h\frac{\sqrt{E^2 - (1-2m/r)}}{E(1-2m/r)} dr. \tag{19}$$

with $d\tau = EdT$. Refs. 8-11 substitute the timelike variable T, or a closely related variable, into the Schwarzschild metric to derive the metric associated with the family of particles on radial Schwarzschild geodesics. This metric, given for the two-dimensional radial geometry in terms of the proper time, is

$$ds^2 = -d\tau^2 + E^{-2}(dr - h[E^2 - (1-2m/r)]^{1/2} d\tau)^2 \tag{21}$$

A similar metric has been derived for particles on radial Reissner-Nordström geodesics[9].

With (14a) the metric given in (21) can be written

$$ds^2 = -d\tau^2 + E^{-2}(dr - \frac{dr}{d\tau} d\tau)^2 \tag{22}$$

The metric is therefore equivalent to $ds^2 = -d\tau^2$ which is the definition of the line element in terms of proper time and is independent of the geometry and the coordinate system. The equations of motion and this universal line element in the form given by (22) can be used to demonstrate that the following elements are not independent for two dimensional radial Schwarzschild geometry expressed in a particular coordinate system:

1. The existence of two families of curves that are everywhere orthogonal, with one family of curves given by the geodesics. This geometry dictates (18).
2. The first order equation of motion for dr/dτ, given by (14a) in Schwarzschild coordinates.
3. The first order equation of motion for dt/dτ, given by (14b) in Schwarzschild coordinates.
4. The two-dimensional Schwarzschild metric, (11).

For example, if elements 1, 2 and 4 are given, the first order equation of motion for dt/dτ can be derived. This interdependence is not surprising. The time coordinate, t, appearing in the metric and in the first order equation of motion for dt/dτ, describes the time measured by a particular observer relative to the proper time of a particle on a geodesic. (The radial motion of the particle is measured by the same observer.)

## V. HAMILTON'S EQUATIONS AND QUANTIZATION

Define the Hamiltonian H by $H = \tfrac{1}{2}(Q^2 + P^2)$ where $Q = E\tanh(2\chi_1)$ and $P = E\,\mathrm{sech}(2\chi_1)$. Since

$$dQ = E\,\mathrm{sech}^2(2\chi_1)d(2\chi_1) = Pd\varepsilon \tag{23a}$$

and

$$dP = -E\tanh(2\chi_1)\mathrm{sech}(2\chi_1)d(2\chi_1) = -Qd\varepsilon \tag{23b}$$

with $d\varepsilon$ defined by (3), Q and P satisfy Hamilton's equations of motion with respect to the time-like variable $\varepsilon$. The Hamiltonian is quadratic in the canonical position and momentum, Q and P, the form of the Hamiltonian for the SU(1,1) harmonic oscillator. Having found this canonical form of the equations of motion from their SU(1,1)/U(1) symmetry, we have at our disposal a more well known quantization procedure[3,4]. In terms of the physical parameters, the Hamiltonian and the canonical position and momentum are given by

$$H = \tfrac{1}{2}E^2, \tag{24a}$$

$$Q = \frac{dr}{d\tau}, \tag{24b}$$

and

$$P = \sqrt{1 - 2m/r}. \tag{24c}$$

## V. DISCUSSION

The equations of motion for any dynamical system with SU(1,1)/U(1) symmetry can, in principle, be formulated to produce Hamilton's equations with a Hamiltonian that is quadratic in the canonical position and momentum.[13] The canonical form of Hamilton's equations have been derived herein from the symmetry group for test particles on a Schwarzschild radial ray. The method used to obtain the appropriate form of the radial equation of motion is the geometry generated by the SU(1,1)/U(1) symmetry. Specifically, a parameterization was chosen so the geodesics of the particles were represented by geodesics within the Poincaré disk model and Poincaré upper half-plan model of hyperbolic geometry, as illustrated in Fig. 1 and Fig. 2. This geometric identification resulted in the determination of a globally defined time-like variable which can be expressed as a function of particle proper time. Hamilton's equations are then established using this variable as the time parameter. Quantization based on the representations of SU(1,1)/U(1) have been extensively studied.[3,4]

One can envision several ways to extend the family of particles discussed herein. For example, one might add the family of particles on all circular geodesic orbits around the Schwarzschild mass that pass through an extremum point of a particle on the radial ray. The relativistic velocity of this particle can be determined relative to the (stationary) particle at the extremum point, as measured by the observer at infinity. Thus, the proper time of the particle with a circular geodesic orbit can be determined.[14, 15] Since the global time parameter of the particles on the radial ray can be related to the proper time, the global time parameter can be extended to include all particles on circular geodesic orbits that intersect the radial ray at an extremum point. Extremum points exist at every point on the radial ray at which circular geodesic orbits exist. Therefore all circular geodesic orbits intersecting the radial ray can be included in the family of particles that share a global time parameter. Also, since these circular geodesic orbits intersect all other radial rays, this process can be continued in reverse to include all particles on all radial rays in the family of particles sharing a global time parameter. Thus, the family of particles on all radial rays and all allowable circular orbits share a global time parameter.

Particle motion on a radial ray has well-defined properties. The symmetry group of the equations of motion is known and simple, being SU(1,1)/U(1); the global time parameter is well defined; the role of the observer and the relationship of the coordinates and the observer are apparent; and the physics of the test mass is well defined. With these attributes the motion of a test mass on a radial ray relative to the Schwarzschild mass should be an illustrative case study within symplectic geometry.

This article analyses the family of particles moving on radial Schwarzschild geodesics along a radial ray. One can also define a family of particles with SU(1,1)/U(1) dynamical symmetry whose geodesics lie in a cone centered on a Kerr black.[16]

**ACKNOWLEDGEMENTS**

The author thanks Jeffrey K. Lawson for his comments on the manuscript.

**Appendix 1**: Projective coordinates for the SU(1,1)/U(1) coset space

A matrix representation of the algebra for SU(2) can be expressed in terms of the well known Dirac matrices, $\sigma_k$, with[5] $su(2) = \sum_{k=1}^{3} \frac{i}{2}\sigma_k \theta^k$. The matrix representation can be given in terms of the decomposition of the generators of the algebra by

$$su(2) = k \oplus p \tag{A1}$$

where k is the algebra for the compact subgroup $S[U(1) \otimes U(1)] \cong U(1)$ of SU(2). The matrix representation of the algebra for the non-compact Lie group SU(1,1) is obtained from this representation of the algebra for the compact group SU(2) by applying the Weyl unitary trick, $p \rightarrow p^*$. $p^*$ is the algebra for su(1,1) mod u(1) corresponding to the group SU(1,1)/U(1), $p^* \Rightarrow SU(1,1)/U(1)$. A matrix representation of SU(1,1)/U(1) is given by

$$p^* \Rightarrow \begin{bmatrix} Y & X \\ \overline{X} & Y \end{bmatrix} = \begin{bmatrix} \cosh(\alpha/2) & (\hat{\alpha}_1 - i\hat{\alpha}_2)\sinh(\alpha/2) \\ (\hat{\alpha}_1 + i\hat{\alpha}_2)\sinh(\alpha/2) & \cosh(\alpha/2) \end{bmatrix} \tag{A2}$$

with $\alpha = \pm\sqrt{\alpha_1^2 + \alpha_2^2}$ and $\hat{\alpha}_i = \alpha_i / \alpha$. A point in the SU(1,1)/U(1) coset space is defined by the homogenous coordinates X and Y, which are defined on the positive sheet of a two-dimensional, two-sheeted hyperboloid. The points in the coset space can also be defined by the projective coordinate z = X/Y. The points in the projective space, z, are the projection of the points on the positive sheet of the two sheeted hyperboloid onto the complex unit disk, the projection being toward the origin of the space containing the two-sheeted hyperboloid, namely X = Y = 0. One can define the points in the unit disk by

$$z = \tanh(\chi)(\hat{x} + i\hat{y}) \tag{A3}$$

with $\hat{x} = \hat{\alpha}_1$, $\hat{y} = -\hat{\alpha}_2$ and $\chi = \alpha/2$ providing the correspondence to the coset space definitions used in (A2).

**Appendix 2**: Continuation outside the unit circle

The arcs of circles within the unit circle in the complex plane that are described in the text can be continued outside the unit circle. As illustrated in Fig. 1, the arcs centered on the y-axis, which are the flow lines of the geodesic transformations, are orthogonal to the unit circle. The arcs can be continued across the unit circle by inversion in the unit circle. A circle is produced by the arc and its inverse. The circle corresponding to the arc that intersects the y-axis at $\tanh(\chi_2)$ has radius equal to $\text{csch}(2\chi_2)$ and is centered at $(0, \coth(2\chi_2))$.

Similarly, the arc in Fig. 1 that is centered on the x-axis is also an arc of a circle, although such arcs are not orthogonal to the unit disk. The circle corresponding to the arc that intersects the x-axis at $\tanh(\chi_1)$ has radius equal to $\coth(2\chi_1)$ and is centered at $(-\cosh(2\chi_1), 0)$.

The arcs and associated circles generated by the flow lines of the geodesic transformation and the extremum transformation, (2), form two families of curves within the Poincaré unit disk and outside the unit disk, and in the entire complex plane associated with the Poincaré upper half-plane model. These families of coaxal curves are everywhere orthogonal. Every point within and outside the unit disk, and in the complex plane, can be defined as the intersection of two unique orthogonal circles, one centered on the x-axis and one centered on the y-axis.